\documentclass[10pt,twocolumn,prl,aps,floatfix,superscriptaddress, longbibliography]{revtex4-1}

\usepackage{color}
\usepackage[color=green!60]{todonotes}
\usepackage{graphicx}
\usepackage{amsthm}
\usepackage{amsmath}
\usepackage{amssymb}
\usepackage{pdfpages}

\makeatletter
\AtBeginDocument{\let\LS@rot\@undefined}
\makeatother

\DeclareMathOperator{\Tr}{Tr}
\newcommand{\ren}{R{\'e}nyi~}

\begin{document}

\title{\ren generalization of the operational entanglement entropy}

\author{Hatem Barghathi}
\affiliation{Department of Physics, University of Vermont, Burlington, VT 05405, USA}

\author{C.~M. Herdman}
\affiliation{Department of Physics,  Middlebury College, Middlebury, VT 05753,
USA}

\author{Adrian Del Maestro}
\affiliation{Department of Physics, University of Vermont, Burlington, VT 05405, USA}
\affiliation{Institut f\"ur Theoretische Physik, Universit\"at Leipzig, D-04103, Leipzig, Germany}

\begin{abstract}
    Operationally accessible entanglement in bipartite systems of indistinguishable particles could be reduced due to restrictions on the allowed local operations as a result of particle number conservation. In order to quantify this effect, Wiseman and Vaccaro [Phys. Rev. Lett. \textbf{91}, 097902 (2003)] introduced an operational measure of the von Neumann entanglement entropy. Motivated by advances in measuring \ren entropies in quantum many-body systems subject to conservation laws, we derive a generalization of the operational entanglement that is both computationally and experimentally accessible.  Using the Widom theorem, we investigate its scaling with the size of a spatial subregion for free fermions and find a logarithmically violated area law scaling, similar to the spatial entanglement entropy, with at most, a double-log leading-order correction. A modification of the correlation matrix method confirms our findings in systems of up to $10^5$ particles.
\end{abstract}
\maketitle

Entanglement encodes the amount of non-classical information shared between complementary parts of an extended quantum state. For a pure state described by density matrix $\rho$, it can be quantified via the \ren entanglement entropies: $S_{\alpha}(\rho_A)=(1-\alpha)^{-1}\ln{\rm{Tr}}\, \rho_A^{\alpha}$ where $\rho_A$ is the reduced density matrix of subsystem $A$ and $S_\alpha$ is a non-increasing function of $\alpha$. While evaluation of the $\alpha=1$ (von Neumann) entanglement entropy requires a complete reconstruction of $\rho$, \cite{Roos:2004hm, Haffner:2005hs}, integer values with $\alpha > 1$ can be represented as the expectation value of a local operator \cite{Calabrese:2004hl}. This has enabled entanglement measurements in a wide variety of many-body states, both via quantum Monte Carlo \cite{Hastings:2010dc, Humeniuk:2012cq, McMinis:2013dp, Herdman:2014ey, Drut:2015fs} and experimental quantum simulators employing ultra-cold atoms \cite{Daley:2012bd, Islam:2015cm, Kaufman:2016ep, Pichler:2016ec, Linke:2017tf, Lukin:2018wg}.  In these systems, conservation of total particle number $N$ may restrict the set of possible local operations, (a superselection rule) and can potentially limit the amount of entanglement that can be physically accessed \cite{Horodecki:2000hr, Bartlett:2003ud, Wiseman:2003jx, Wiseman:2003vn, Vaccaro:2003ei, Schuch:2004jv,Dunningham:2005fu,Cramer:2011cx}.  For example, while a superfluid of $N$ bosonic $^{87}$Rb atoms in a one-dimensional optical lattice is highly entangled under a bipartition into spatial subregions \cite{Islam:2015cm}, much of the entanglement is generated by particle fluctuations that cannot be transferred to a quantum register without access to a global phase reference \cite{Aharonov:1967be}.  Wiseman and Vaccaro introduced  an \emph{operational} measure of entropy to quantify these effects \cite{Wiseman:2003jx}, but it is limited to the special case of $\alpha=1$ and thus cannot be used in tandem with current simulation and experimental studies of entanglement.

In this paper, we study how the operational entanglement can be generalized to the \ren entropies with $\alpha  \ne 1$.  Recalling its definition for $\alpha=1$, it is constructed by averaging the contributions to $S_1$ coming from each physical number of particles in the subsystem:
\begin{equation}
S_1^{\rm{op}}(\rho_A)=\sum_{n=0}^{N} P_n S_1(\rho_{A_n})
\label{eq:S1op}
\end{equation}
where $\rho_{A_{n}} = \mathcal{P}_{A_{n}} \rho_A \mathcal{P}_{A_{n}}/P_n$ is the  
projection into the sector of $n$ particles in $A$, $A_n$, via
$\mathcal{P}_{A_{n}}$ which occurs with probability $P_n={\rm{Tr}}\,
\mathcal{P}_{A_{n}} \rho_A \mathcal{P}_{A_{n}}$. This projection constitutes a local operation which can only decrease entanglement by an amount bounded by the maximum entropy of the classical number fluctuation probability distribution $P_n$.  Thus, a conservation law on the total number of particles imposes that any \ren generalization of Eq.~(\ref{eq:S1op}) to $S_\alpha^{\rm op}$ must satisfy $0 \le S_\alpha - S_\alpha^{\rm op} \le \ln D$ where $D$ is the support of $P_n$.  Under this physical constraint, we show that 
a direct extension of Eq.~(\ref{eq:S1op}) to $\alpha \ne 1$ is not generally appropriate. 

Instead, we reconsider the problem in terms of the mathematical relationship between the von Neumann and $\alpha\ne1$ \ren entropies -- that of a geometric to power mean -- and identify a unique measure:
\begin{equation}
S_{\alpha}^{{\rm op}} (\rho_A) = \frac{\alpha}{1-\alpha}\ln\,  \sum_{n} P_n \mathrm{e}^{\frac{1-\alpha}{\alpha} S_{\alpha}(\rho_{A_n})}
\label{eq:Saop}
\end{equation}
which not only provides a lower bound on the amount of operational entanglement entropy in a pure state, but is accessible with current technologies for integer $\alpha > 1$.

We validate that Eq.~(\ref{eq:Saop}) reproduces Eq.~(\ref{eq:S1op}) as $\alpha \to 1$ and prove that it is a non-increasing function of \ren index $\alpha$ in analogy with $S_\alpha $.  We show that $S_\alpha^{\rm op} = 0$ when all particles have condensed into a single mode, \emph{e.g.}~a Bose-Einstein condensate, and demonstrate that in the limit of large subsystem size, it agrees with the known behavior of $S_1^{\rm op}$ for free fermions in $d$ spatial dimensions \cite{KlichLevitov:2008} -- that the fixed total particle number reduces the operational entanglement only by a subleading logarithm, $S_\alpha^{\rm op} \approx S_\alpha - \tfrac{1}{2}\ln S_\alpha$.  Such asymptotic scaling is expected for $1d$ critical systems with fixed $N$ that can be described by a conformal field theory, where the particle number distribution is Gaussian \cite{Song:2010eq, Song:2012cp}.  

The main contributions of this work are (1) the introduction of the \ren generalization of the operational entanglement entropy; (2) an investigation of its asymptotic scaling properties for free fermions via the Widom theorem supported by exact calculations for non-interacting $1d$ lattice fermions; and (3) a discussion of how the operational entanglement could be measured in ultra-cold atomic lattice gases using current technology.


We begin by recognizing that the von Neumann  entanglement entropy $S_{1}(\rho_A)=-{\rm{Tr}}\, \rho_A\ln \rho_A$ can be written as the negative logarithm of the geometric mean $\mathfrak{s}_{1}(\rho_A) \equiv \exp[-S_{1}(\rho_A)]={\rm{det}}\, \rho_A^{\rho_A}$ which is the mean of $\rho_A$ over $\rho_A$. The \ren entanglement entropies are then obtained by generalizing the geometric mean $\mathfrak{s}_{1}(\rho_A)$ to the power mean: $\mathfrak{s}_{\alpha}(\rho_A)=\left({\rm{Tr}}\, \rho_A\rho_A^{\alpha-1}\right)^{(\alpha-1)^{-1}}$. With this in mind, we rearrange the expression for $S_1^{\rm{op}}(\rho_A)$ in Eq.~(\ref{eq:S1op}) as $S_1^{\rm{op}}(\rho_A)=-\ln\Pi_n\left[\mathfrak{s}_{1}(\rho_{A_n})\right]^{P_n}$, which is  the negative logarithm of the  geometric mean of $\mathfrak{s}_{1}(\rho_{A_n})$ over the distribution $P_n$. Thus we can obtain a \ren generalization of $S_1^{\rm{op}}(\rho_A)$ by replacing the geometric mean $\mathfrak{s}_{1}(\rho_{A_n})$ with power mean $\mathfrak{s}_{\alpha}(\rho_{A_n})$ and the geometric mean over $P_n$ with a power mean of order $\gamma$: $S_{\alpha}^{\rm{op}}(\rho_A;\gamma)=-\ln\left[\sum_n P_n \mathfrak{s}_{\alpha}(\rho_{A_n})^{\gamma}\right]^{\gamma^{-1}}$  where $\gamma = \gamma(\alpha)$ is yet to be determined. In the limit $\gamma\to 0$, one recovers the direct extension of Eq.~(\ref{eq:S1op}): $S_{\alpha}^{\rm{op}}(\rho_A;0)=\sum_{n=0}^{N} P_n S_{\alpha}(\rho_{A_n})$  which was previously proposed to study a system of bosons in one dimension \cite{Melko:2016bo}. 

Defining $\Delta S_\alpha(\gamma) \equiv S_{\alpha}(\gamma) - S_{\alpha}^{\rm{op}}(\gamma)$, we now explore what restrictions are imposed on the exponent $\gamma$ by the physical constraint that $0 \le\Delta S_\alpha(\gamma)\le \ln D$.  
To this end, we consider the example of a reduced density matrix of a spatial partition of $\ell$ sites, obtained from a pure state of $N\gg1$ particles, where the number fluctuations are described by the normalized distribution: $P_n=A_N\exp[-(N-n)/{\sqrt{N}}]$.  The corresponding eigenvalues of $\rho_A$ are equal for each $n$: $\lambda_{n,i}=\ell^{-n}A_N\exp[-(N-n)/{\sqrt{N}}]$ where $i = 1,\ldots,\ell^n$.  In this case, $D=N+1$ and the asymptotic dependence  of $\Delta S_{\alpha>1}(\gamma)$, to leading order, on $N$ for $\gamma\neq 1-\alpha^{-1}$ is given by $\Delta S_{\alpha>1}(\gamma) \approx (\tfrac{\alpha}{\alpha-1}-\tfrac{1}{\gamma})\sqrt{N}$ for $\gamma>0$ and $\Delta S_{\alpha>1}(\gamma) \approx -N\ln\ell$ for $\gamma\le 0$ which violates the condition $0 \le\Delta S_\alpha(\gamma)\le \ln D$ for any $\gamma\neq 1-\alpha^{-1}$. If we modify the above example by rearranging the probabilities in the reverse order, \emph{i.e.}~replacing $P_n$ with $P_{N-n}$, we arrive at the same conclusion for $\alpha<1$ (see supplemental material \cite{supplemental} for complete proof.)

For $\gamma = 1-\alpha^{-1}$ we define $x_n=P_n^\alpha\,{\rm{Tr}}\, \rho_{A_n}^{\alpha}$ and can write $\mathrm{e}^{\Delta S_\alpha(1-\alpha^{-1})}=\left(\| X\|_{\alpha^{-1}}/\| X\|_{1}\right)^{1/(\alpha-1)}$, where, $\| X\|_p=\left(\sum_n \vert x_n\vert^p\right)^{p^{-1}}$ is the $p$-norm of the vector $X=\{x_n\}$. The property $\| X\|_q\le\| X\|_r\le D^{r^{-1}-q^{-1}}\| X\|_q$ holds for $0<r\le q$, $q,r \in \mathbb{R}$, guaranteeing that $0 \le\Delta S_\alpha(\gamma)\le \ln D$ is satisfied for $\gamma=1-\alpha^{-1}$. For this power mean exponent, it can also be shown that $S_{\alpha}^{\rm{op}}$ is a lower bound for $S_1^{\rm{op}}$ for $\alpha > 1$ (upper bound for $\alpha < 1$), \emph{i.e.}~$S_{\alpha}^{\rm{op}}$ is a non-increasing function of $\alpha$, and by construction, $\lim_{\alpha\to 1} S_\alpha^{\rm op} = S_1^{\rm op}$ \cite{supplemental}. Thus we propose Eq.~(\ref{eq:Saop}) as the unique \ren generalization of the operational entanglement entropy.

For more physical insight into the form of this measure, we appeal to a previously noticed connection between the von Neumann operational entanglement and the Shannon conditional entropy \cite{Horodecki:2005fv, KlichLevitov:2008}.  If the spectrum of the reduced density matrix $\rho_A$ is treated as a joint probability distribution of two random variables, one of which is the number of particles $n$ in partition $A$, then Eq.~(\ref{eq:S1op}) is equivalent to the conditional entropy of the probability distribution, where the condition is information of $n$ in the subregion.  Many different candidate measures for the classical conditional \ren entropy have been proposed \cite{Cachin97entropymeasures,GolshaniPashaYari:2009,Hayashi:2011,SKORIC:2011el,FehrBerens2014}, but if one requires that they satisfy both monotonicity and the weak chain rule, then the classical limit of Eq.~(\ref{eq:Saop}) is recovered.

Having understood the origin of the \ren generalized operational entanglement entropy, in order to actually perform computations, we exploit that fact that for pure states of $N$ particles, $\rho_A$ is block diagonal in $n$ and thus Eq.~(\ref{eq:Saop}) can be conveniently rewritten as
%
%
%
\begin{equation}
S_{\alpha}^{{\rm op}} = S_{\alpha}-H_{1/\alpha}\left(\{P_{n,\alpha}\}\right)
\label{eq:S_alpha_op5}
\end{equation}
where $H_{\alpha}\left(\{P_n\}\right)=(1-\alpha)^{-1}\ln\sum_n P_n^{\alpha}$ is the \ren generalization of the Shannon entropy of $P_n$, 
\begin{equation}
    P_{n,\alpha}=\frac{ \Tr\, \left[\mathcal{P}_{A_{n}} \rho_A^{\alpha} \mathcal{P}_{A_{n}}\right]}{\Tr\, \rho_A^{\alpha}}
\label{eq:Pna}
\end{equation}
is a normalization of partial traces of $\rho_A^{\alpha}$, and $P_{n,1}=P_n$. From Eq.~(\ref{eq:S_alpha_op5}) one immediately recovers the previously known result for $\alpha=1$ that $\Delta S_1 = H_1$ \cite{KlichLevitov:2008} where we write $H_\alpha \equiv H_\alpha(\{P_n\})$ for simplicity. 

In the remainder of this paper we use Eqs.~(\ref{eq:S_alpha_op5}) and (\ref{eq:Pna}) to calculate the \ren generalized operational entanglement for two simple models of non-interacting particles.  First, we consider the case of $N$ non-interacting bosons on a $d$-dimensional hypercubic lattice of $L^d$ sites with unit lattice spacing.  The ground state consists of all particles condensed into one single-particle mode $\vert \Psi \rangle=(N!)^{-1/2}(\Phi^\dag_0)^N\vert0\rangle$ where $\Phi_0^\dag = \sum_j B_j b_j^\dag$ and $b_j^\dag$ creates a boson on site $j$ with $\sum_j|B_j|^2=1$.  We take a spatial bipartition $A$ that contains a set of $\ell^d$ contiguous sites and decompose $\Phi^\dag_0=\sqrt{p_{A}}\Phi^\dag_{A} +\sqrt{p_{\bar{A}}}\Phi^\dag_{\bar{A}}$ with $p_A=\vert\langle 0\vert\Phi_{A}\Phi^\dag_{0}\vert0\rangle\vert^2$, $p_{\bar{A}}=1-p_A$ and $\Phi^\dag_{A}$ acts in $A$, similarly for the complement $\bar{A}$. Then, the ground state can be directly written as the Schmidt decomposition  
\begin{equation*}
    \vert \Psi\rangle=\sum_{n=0}^N\lambda_n^{1/2}\; \vert n\rangle_{A} \otimes \vert N-n\rangle_{\bar{A}}
\end{equation*}
where $\lambda_n = {{N}\choose{n}}p_A^np_{\bar{A}}^{N-n}$, $\vert n\rangle_{A}=(n!)^{-1/2}(\Phi^\dag_A)^n\vert0\rangle_A$ and $\vert N-n\rangle_{\bar{A}}=[(N-n)!]^{-1/2}(\Phi^\dag_{\bar{A}})^{N-n}\vert0\rangle_{\bar{A}}$. For free bosons $p_A=\left(\ell /L\right)^d$  \cite{Simon:2002it, Herdman:2014ey}.  The reduced density matrix $\rho_A$ obtained by tracing out $\bar{A}$ is thus pure for each $n$: $\rho_{A_{n}} = \vert n\rangle{\langle}n\vert$ resulting in $S_\alpha=H_\alpha$ and $P_{n,\alpha} = P_{n}^\alpha / \sum_n P_n^\alpha \Rightarrow S_\alpha^{\rm op} = 0$. This is expected for the Bose-Einstein condensate where for $N \gg 1$ with $p_A$ fixed, $P_n = \lambda_n$ approaches a Gaussian distribution and $S_\alpha=H_\alpha \approx \tfrac{1}{2}\ln N$ \cite{Simon:2002it,Ding:2009gq} is generated from particle fluctuations between subregions.

To understand the behavior of $S_\alpha^{\rm op}$ for fermionic statistics, we focus on a microscopic model of non-interacting fermions on a $d$-dimensional lattice where the correlation matrix method \cite{Peschel:2003,Peschel:2009,EislerPeschel:2017,Peschel:2012wq,CalabreseLefevre:2008} is applicable. This provides an exponential simplification of the calculation of $S_{\alpha}(\rho_{A})$ and allows for the investigation of its asymptotic behavior.  In this case, $A$ corresponds to some collection of $\ell^d$ lattice sites and the eigenvalues of $\rho_A$ that correspond to having $n$ particles in partition $A$, are $\lambda_{n,a}=\prod_{j=1}^{\ell^d}\left[{\nu_j^{n_{j,a}}\bar{\nu}_j^{(1-n_{j,a})}}\right]$, where the index $a$ runs over all possible configurations of the occupation numbers $n_{j,a} \in \{0,1\}$ with $n =\sum_j n_{j,a} \; \forall a$  and $\bar{\nu}_j=1-\nu_j$. Here, $\nu_j$ are the eigenvalues of the correlation matrix $(\mathsf{C}_A)_{ij}=\langle c^\dagger_{i} c^{\phantom{\dagger}}_{j}\rangle={\Tr}\, \rho_A c^\dagger_{i} c_j^{\phantom{\dagger}}$ where $i,j$ are restricted to the spatial partition $A$  and  $c_i^\dag(c_i)$ creates (annihilates) a spinless fermion at lattice site $i$ ($c_i^{\phantom \dag}c^\dag_j + c^\dag_j c_i^{\phantom \dag} = \delta_{ij}$) \cite{Peschel:2003}. 

This approach can be generalized to calculate the particle number projected \ren entanglement $S_{\alpha}(\rho_{A_n})=S_{\alpha}+{(1-\alpha)}^{-1}\ln\left(P_{n,\alpha}/P_n^{\alpha}\right)$ and thus  $S_\alpha^{\rm op}(\rho_{A})$. However, as we are interested in the reduction of entanglement due to the presence of superselection rules, we focus on the difference $\Delta S_{\alpha}= S_{\alpha}-S_{\alpha}^{{\rm op}}$ which depends only on:
\begin{equation}
P_{n,\alpha}=\sum_a\prod_{j=1}^{\ell^d}\left[{\nu_{j,\alpha}^{n_{j,a}}\bar{\nu}_{j,\alpha}^{(1-n_{j,a})}}\right]
\label{eq:P_n_alpha},
\end{equation}
where $\nu_{j,\alpha}=\nu_j^{\alpha}/(\nu_j^{\alpha}+\bar{\nu}_j^{\alpha})$. An important first step is the observation that $P_{n,\alpha}$ has the form of a Poisson-binomial distribution \cite{PoissonBinomial} with $\ell^d$ different success probabilities $\nu_{j,\alpha}$ \footnote{The mean and the variance of the Poisson-binomial distribution are given by $\sum_j\nu_j $ and  $\sum_j\nu_j\bar{\nu}_j $, respectively.}.  In order to investigate the asymptotic scaling of $\Delta S_{\alpha}$ with linear subsystem size $\ell$ we need to consider the behavior of $P_{n,\alpha}$  or, alternatively, its  characteristic function (Fourier transform) $\chi_{\alpha}(\lambda)=\prod_{i=1}^{\ell^d}\left[1-\nu_{j,\alpha}+\nu_{j,\alpha} ~{\rm e}^{i\lambda}\right]$ which can be expressed in terms of the matrix ${\sf C}_A$ as
\begin{equation}
    \ln\chi_{\alpha}(\lambda)={\Tr}\,  \ln\left[1-{\sf C}_{A,\alpha} +{{\sf C}_{A,\alpha} } ~{\rm e}^{i\lambda}\right]
\label{eq:Chi_alpha},
\end{equation}
where ${\sf C}_{A,\alpha} \equiv {\sf C}_A^\alpha/[{\sf C}_A^\alpha+(1-{\sf
C}_A)^\alpha]$. This form is convenient, as the $\alpha=1$ case, providing
access to the scaling of $P_{n,1} = P_n$, has already been obtained for the
$d$-dimensional free Fermi gas by means of the Widom theorem
\cite{LandauWidom1980,Widom1982, GioevKlich:2006,
KlichLevitov:2008,Sobolev:2013, Leschke:2014el, Sobolev:2014, Sobolev2015}.
Motivated by these results, we calculate the characteristic function
$\chi_{\alpha}(\lambda)$ for a $d$-dimensional spatial subregion with
dimensionless linear size $\ell$ in the limit $\ell\gg 1$ where $\ell$ is now
treated as a continuous variable. We find that, $P_{n,\alpha}$ is a normal distribution with the  same average as $P_n$ and  variance $\sigma^2_{\alpha}=\sigma^2/\alpha\sim \ell^{d-1}\ln{\ell}/\alpha$, where $\sigma^2$ is the variance of $P_{n}$ \cite{supplemental}. In this case, $P_{n,\alpha}\sim P_n^\alpha \Rightarrow H_{1/\alpha}(\{P_{n,\alpha}\})=H_{\alpha}(\{P_n\})$ leading to
\begin{equation}
    \Delta S_{\alpha}\approx H_{\alpha}\approx\ln \sqrt{2\pi\sigma^2\alpha^{1/(\alpha-1)}}\sim \tfrac{1}{2} \ln\left(\ell^{d-1}\ln{\ell}\right)
\label{eq:Ds_alpha_asymptotic},
\end{equation}
which, if compared to the asymptotic scaling of  $S_{\alpha}\sim \ell^{d-1}\ln{\ell}$ \cite{Leschke:2014el}, implies that $\Delta S_{\alpha}\approx \tfrac{1}{2}\ln S_{\alpha}$. We thus conclude that fixed $N$ only reduces the \ren generalized operational entanglement of the free Fermi gas by a subleading double logarithm of $\ell$ for $\ell \gg 1$.

To confirm the asymptotic predictions of Eq.~(\ref{eq:Ds_alpha_asymptotic}) we now apply the extended correlation matrix method introduced above to a model of $N$ free spinless lattice fermions on a ring of $2N$ sites (half-filling) governed by the Hamiltonian $\mathcal{H} = -\sum_{i}(c^\dagger_{i} c^{\phantom{\dagger}}_{i+1} + \text{h.c.})$ \cite{repo}. The correlation matrix for the ground state Fermi sea is $({\sf C}_A)_{ij}=\frac{\sin[\pi(i-j)/2]}{2N\sin[\pi(i-j)/2N]}$. We studied systems with up to $N=10^5$ fermions and partition sizes $\ell=10^5$ sites, where we calculate $\Delta S_{\alpha}$ and $H_{\alpha}$ using $P_{n,\alpha}$ which we obtain via a recursion relation for the Poisson-binomial distribution \cite{BarlowHeidtmann:1984}:
\begin{equation}
P_{n,\alpha}(j)=\nu_{j,\alpha} P_{n-1,\alpha}(j-1)+\bar{\nu}_{j,\alpha}P_{n,\alpha}(j-1).
\label{eq:P_n_recursion}
\end{equation}
The desired distribution is reached after $\ell$ recursive steps, \emph{i.e.}~$P_{n,\alpha}=P_{n,\alpha}(\ell)$ and Eq.~(\ref{eq:P_n_recursion}) drastically reduces the complexity to an $\mathrm{O}(\ell^2)$ algorithm \cite{BarlowHeidtmann:1984}.

%
\begin{figure}[t]
\begin{center}
\includegraphics[width=1.0\columnwidth]{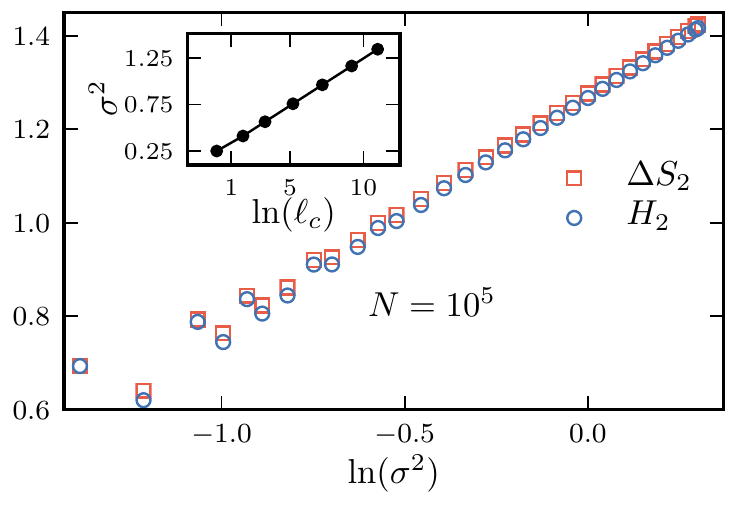} 
\end{center}
\caption{Scaling of the difference between the \ren and operational entanglement entropy, $\Delta S_{2}$ and $H_{2}$, with the log of the variance of $P_n$, $\ln(\sigma^2)$, for subregions up to $\ell=10^5$ connected sites.  The results were calculated using the correlation matrix method for free fermions in the ground state of $\mathcal{H}$.  Inset: Scaling of $\sigma^2$ with $\ln(\ell_c)$, where $\ell_c=(2N/\pi)\sin[\pi\ell/(2N)]$ is the chord length, highlighting the double logarithmic growth of the width of the distribution $P_n$.} 
\label{FiniteSizeScalingofDS2}
\end{figure}
\begin{figure}[t]
\begin{center}
\includegraphics[width=1.0\columnwidth]{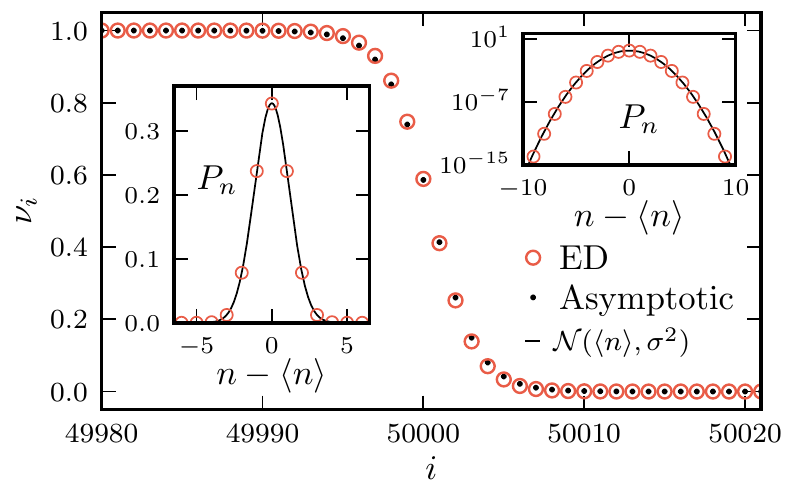} 
\end{center}
\caption{The spectrum of the correlation matrix ${\sf C}_A$ of free fermions calculated via exact diagonalization (empty circles) and from the asymptotic relation in Eq.~(\ref{eq:nu_asymptotic}) (filled circles) for $N=10^5$ at half-filling with partition size $\ell=10^5$.  Insets: The corresponding number probability distribution $P_n$ vs $n-\langle n \rangle$ on a linear (left) and log (right) scale.  The solid line shows a normal distribution $\mathcal{N}$ with the average $\langle n \rangle$ and variance $\sigma^2$ of $P_n$ demonstrating its convergence but narrow width.}
\label{AsymptoticEigenvalues}
\end{figure}

%
The results in Fig.~\ref{FiniteSizeScalingofDS2} demonstrate the predicted logarithmic scaling of $\Delta S_{2}$ with $\sigma^2=2\sigma^2_2$ as well as the fact that asymptotically, $\Delta S_{2}\approx H_{2}$, \emph{i.e.}~that $P_{n,2}$ appears to behave as a continuous normal distribution.  For this particular case of free fermions we find that $S_\alpha - S_\alpha^{\rm op} > H_\alpha$, but this may not be generically true in interacting models. Additionally, as seen in Fig.~\ref{AsymptoticEigenvalues}, $P_n$ is very narrow, with $\sigma^2<1.4$ and thus the main contribution comes from only a few points around its peak.  This suggests that to truly reach the asymptotic regime, we need to further increase $\sigma^2$ by several orders of magnitudes beyond our current numerical capability.
%
%

\begin{figure}[th!]
\begin{center}
\includegraphics[width=1.0\columnwidth]{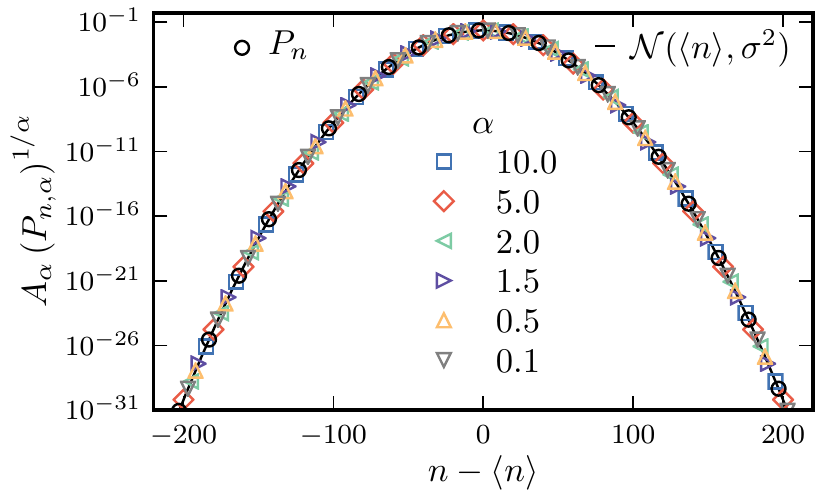} 
\end{center}
\caption{Collapse of the rescaled probability distribution $A_\alpha (P_{n,\alpha})^{1/\alpha}$ to $P_n$ for different values of $\alpha$, where $A_{\alpha}$ is a normalization factor. The solid line shows a normal distribution $\mathcal{N}$ with the average $\langle n \rangle$ and variance $\sigma^2$ of $P_n$. The data was obtained using the correlation matrix method with the asymptotic eigenvalues $\nu_j$ (Eq.~(\ref{eq:nu_asymptotic})) and $\ln \ell =3000$. We find perfect collapse for both integer (supported by the Widom Theorem) and non-integer values of $\alpha$.}
\label{CollapsedDistributions}
\end{figure}

As an alternative, we generalize the known asymptotic behavior of $\nu_j$ \cite{EislerPeschel:2013,Slepian:1978,Peschel:2004} to $\nu_{j,\alpha}$ as
\begin{equation}
    \nu_{j,\alpha}= \left[1+\exp\left(\frac{-\alpha\pi^2(\ell-2j+1)}{2[\ln(8\ell)+\gamma_{\rm em}]}\right)\right]^{-1}
  \label{eq:nu_asymptotic},
\end{equation}
where $\gamma_{\rm em} \approx 0.6$ is the Euler–Mascheroni constant and calculate the characteristic function $\chi_{\alpha}(\lambda)$ of $P_{n,\alpha}$. We find that $P_{n,\alpha}$ is asymptotically a normal distribution with variance $\sigma^2_{\alpha}=\ln\ell / (\alpha\pi^2)$ for any $\alpha>0$ \cite{supplemental} extending the results of the Widom Theorem for $d=1$ to real valued $\alpha$. This is further validated using Eq.~(\ref{eq:nu_asymptotic}) with $\ell\approx \mathrm{e}^{3000}$ as shown in Fig.~\ref{CollapsedDistributions}.
%
%

Thus for free fermions, superselection rules fixing the total number of particles only marginally reduce the operational entanglement that can be transferred from a many-body state to a quantum register. This is also true for interacting $1d$ fermions in the Luttinger liquid regime \cite{KlichLevitov:2008, Goldstein:2018kf}.  The free fermion result is robust even when extending to non-contiguous subregions, \emph{e.g.}~a partition of size $\ell=N$ corresponding to even (odd) sites where the correlation matrix is diagonal and $\nu_{j,\alpha}=\nu_ j=\tfrac{1}{2}$.  Here, $S_{\alpha} = \ell \ln 2$ and $P_{n,\alpha}=P_n,\ \forall \alpha$  are described by a simple Binomial distribution (normal distribution, asymptotically) with $\ell$ equal success probabilities $\nu=\tfrac{1}{2}$.  Thus, $\sigma^2=\ell/4$ and $\Delta S_{\alpha}\sim \ln\sigma^2$ yielding $\Delta S_{\alpha}\sim \frac{1}{2}\ln S_{\alpha}$.  

This picture can be drastically altered by strong interactions \cite{BarghathiCasiano-DiazDelMaestro:2018} or in bosonic systems \cite{Melko:2016bo}, where the contribution of particle fluctuations to entanglement are large and the operational entanglement is suppressed to zero.  

In summary, by exploiting a general relation between geometric and power means, we derive a unique measure $S_\alpha^{\rm op}$ in Eq.~(\ref{eq:Saop}) which generalizes the operational entanglement in the presence of a superselection rule, previously defined only for von Neumann entropies, to the more readily measurable \ren entanglement entropies $S_\alpha$.

This definition preserves the limit $\alpha \to 1$, provides a lower bound on $S_1^{\rm op}$ for $\alpha > 1$, and is smaller than $S_\alpha$ while not exceeding the maximum information lost to particle fluctuations.  $S_\alpha^{\rm op} = 0$ for a Bose-Einstein condensate of fixed total particle number, while for free fermions, we find that the corresponding superselection rule reduces the amount of operational entanglement from its unconstrained value by a subleading correction that asymptotically scales as the logarithm of the width of the probability distribution describing particle fluctuations in the subregion.  We confirm this prediction numerically using the correlation matrix method on a lattice model of free fermions, where we have simplified the calculation by relating the required partial traces $\rho_A^{\alpha}$ to the Poisson-binomial distribution which can be calculated using a simple recursion relation.  This method can be extended to other models of non-interacting fermions, including those with long-range or correlated hopping as well as disordered systems, where contributions to the entanglement entropy from particle fluctuations will be further suppressed.  It is interesting to speculate on how the ideas discussed here could be further generalized to understand the effects of superselection rules on entanglement without resorting to a particular mode bipartition \cite{BarnumHoward:el:Viola2004, ZanardiLidarSeth:2004, BenattiFloreaniniMarzolino:2010, BenattiFloreaniniMarzolino:2014}.

The functional form of the \ren generalized operational entanglement depends only on the full and particle number projected reduced density matrices that can be directly computed by creating copies of a physical system.  It is thus accessible using current simulation \cite{Hastings:2010dc, Humeniuk:2012cq, McMinis:2013dp, Herdman:2014ey, Drut:2015fs} and experimental \cite{Islam:2015cm,Linke:2017tf,Lukin:2018wg} techniques for both bosons and fermions for integer $\alpha \ge 2$ by histogramming $\rho_A^\alpha$ into bins corresponding to the number of particles $n$ observed in the subregion with appropriate post-selection \cite{Melko:2016bo}.  The experimental measurement of the \ren generalized operational entanglement entropy and confirmation of its robust scaling in fermionic systems would, in combination with a protocol for its extraction and transfer to a register, support such many-body phases as a potential resource for quantum information processing.\\[0.5ex]

We would like to thank P.N.~Roy, R.G.~Melko and I.~Klich for insightful discussions and E.~Sela for directing our attention to Ref.~[\onlinecite{Goldstein:2018kf}].  This research was supported in part by the National Science Foundation (NSF) under Award No.~DMR-1553991 and a portion was performed at the Aspen Center for Physics, which is supported by NSF grant PHY-1607611. A.D. acknowledges the German Science Foundation (DFG) for financial support via grant RO 2247/10-1.

\bibliography{refs}




\section*{Supplementary material}

# Supplementary material for "Rényi generalization of the operational entanglement entropy"

Hatem Barghathi, C. M. Herdman and Adrian Del Maestro

In this supplement we consider a proposed measure $S_\alpha^{\mathrm{op}}(\rho_A;\gamma)$ generalizing Wiseman and Vaccaro's [S1] operational entanglement entropy $S_1^{\mathrm{op}}(\rho_A)$ to Rényi index $\alpha \neq 1$ up to an undetermined exponent $\gamma(\alpha)$ and prove that the proposed measure can satisfy a set of minimal necessary physical requirements on information quantities if and only if $\gamma = (\alpha-1)/\alpha$.

We also include additional details on how the measure scales asymptotically with spatial subregion size for a model of non-interacting fermions in $d$-dimensions and free lattice fermions in 1-dimension.

## PHYSICAL CONSTRAINTS ON OPERATIONAL ENTANGLEMENT MEASURES

We propose three necessary conditions that any measure quantifying the operational entanglement $S_\alpha^{\mathrm{op}}$ must satisfy:

(1) $S_\alpha^{\mathrm{op}}$ cannot exceed the Rényi entanglement entropy $S_\alpha$ of the state as the amount of accessible information cannot increase when constrained by superselection rules and the difference $\Delta S_\alpha = S_\alpha - S_\alpha^{\mathrm{op}}$ cannot exceed the maximum amount of information $\ln D$ that can be lost to particle number fluctuations, where $D$ is the support of the particle number probability distribution $P_n$, i.e. $0 \leq \Delta S_\alpha \leq \ln D$.

(2) $S_\alpha^{\mathrm{op}}$ is a non-increasing function of $\alpha$ and can thus provide a lower bound on $S_1^{\mathrm{op}}$ for $\alpha > 1$ and an upper bound on $S_1^{\mathrm{op}}$ for $\alpha < 1$.

(3) $S_\alpha^{\mathrm{op}}$ must reproduce Wiseman and Vaccaro's definition of the operational entanglement entropy as $\alpha \to 1$.

## FIXING THE POWER MEAN EXPONENT $\gamma$

Consider the reduced density matrix $\rho_A$, corresponding to a spatial partition $A$, that is obtained from the state $|\Psi\rangle$ describing some quantum system of a fixed number of particles $N$. Fixing the number of particle in the system guarantees that $[\rho_A, \hat{n}] = 0$, where $\hat{n}$ is the particle number operator acting in $A$, and thus $\rho_A$ is block diagonal in $n$. Therefore, we can write $\rho_A^\alpha = \sum_n (\mathcal{P}_{A_n} \rho_A \mathcal{P}_{A_n})^\alpha$, with $\mathcal{P}_{A_n}$ the projection into the sector $A_n$ of $n$ particles in $A$. Number fluctuations in $A$ are described by the distribution $P_n = \mathrm{Tr}\, \mathcal{P}_{A_n} \rho_A \mathcal{P}_{A_n}$ and the proposed generalized measure of the operational entanglement takes the form:

$$S_\alpha^{\mathrm{op}}(\rho_A;\gamma) = -\frac{1}{\gamma} \ln \left[ \sum_n P_n e^{-\gamma S_\alpha(\rho_{A_n})} \right], \tag{S1}$$

where $\rho_{A_n} = \mathcal{P}_{A_n} \rho_A \mathcal{P}_{A_n}/P_n$ is the projected reduced density matrix defined in the main text along with

$$S_\alpha(\rho_A) = \frac{1}{1-\alpha} \ln (\mathrm{Tr}\, \rho_A^\alpha) \qquad \text{Rényi entropy} \tag{S2}$$

$$H_\alpha = \frac{1}{1-\alpha} \ln \left( \sum_n P_n \right) \qquad \text{Shannon information}. \tag{S3}$$

Next, we consider two examples which lead us to conclude that if $\gamma \neq (\alpha-1)/\alpha$, then condition (1) cannot be satisfied in general.

**Example 1.** Consider the reduced density matrix of a spatial partition of $\ell$ sites, obtained from a pure state of $N \gg 1$ particles, where the number fluctuations are described by the exponential distribution: $P_n = A_N \exp[-(N-n)/\sqrt{N}]$, where $A_N \approx N^{-1/2}$ and $D = N + 1$. The corresponding eigenvalues of $\rho_A$ are equal for each $n$: $\lambda_{n,i} = \ell^{-n} A_N \exp[-(N-n)/\sqrt{N}]$ where $i = 1, \ldots, \ell^n$. In this case

$$S_{\alpha>1}(\rho_A) = \frac{\alpha}{\alpha-1}\sqrt{N} + \frac{\alpha}{\alpha-1}\ln\sqrt{N} + \frac{1}{\alpha-1}\ln(1-\ell^{1-\alpha}) + \mathcal{O}\left(N^{-\frac{1}{2}}\right), \tag{S4}$$

and

$$S^{\text{op}}_{\alpha>1}(\rho_A;\gamma) = \begin{cases} \frac{1}{\gamma}\sqrt{N} + \frac{1}{\gamma}\ln\sqrt{N} + \frac{1}{\gamma}\ln(1-\ell^{-\gamma}) + \mathcal{O}\left(N^{-\frac{1}{2}}\right) & ;\ \gamma > 0 \\ (N - \sqrt{N} + \frac{1}{2})\ln\ell + \mathcal{O}\left(N^{-\frac{1}{2}}\right) & ;\ \gamma = 0 \\ N\ln\ell + \frac{1}{\gamma}\ln\sqrt{N} + \frac{1}{\gamma}\ln(1-\ell^{\gamma}) + \mathcal{O}\left(N^{-\frac{1}{2}}\right) & ;\ \gamma < 0 \end{cases} \quad \text{(S5)}$$

Accordingly, to leading order in $N$

$$\Delta S_{\alpha>1}(\gamma) \approx \begin{cases} \left(\frac{\alpha}{\alpha-1} - \frac{1}{\gamma}\right)\sqrt{N} & ;\ \gamma > 0 \\ -N\ln\ell & ;\ \gamma \leq 0 \end{cases}. \quad \text{(S6)}$$

Therefore, for $\alpha > 1$, Condition **(1)** is violated for any $\gamma \neq \frac{\alpha-1}{\alpha}$.

**Example 2.** Here we modify Example 1 by rearranging the probabilities in the reverse order, *i.e.* $P_n = A_N \exp[-n/\sqrt{N}]$ and $\lambda_{n,i} = \ell^{-n} A_N \exp[-n/\sqrt{N}]$ with $i = 1,\ldots,\ell^n$, where $A_N \approx N^{-1/2}$ and $D = N + 1$. It follows that

$$S_{\alpha<1}(\rho_A) = N\ln\ell + \frac{\alpha}{\alpha-1}\sqrt{N} + \frac{\alpha}{\alpha-1}\ln\sqrt{N} + \frac{1}{\alpha-1}\ln(1-\ell^{\alpha-1}) + \mathcal{O}\left(N^{-\frac{1}{2}}\right), \quad \text{(S7)}$$

and

$$S^{\text{op}}_{\alpha<1}(\rho_A;\gamma) = \begin{cases} \frac{1}{\gamma}\ln\sqrt{N} + \frac{1}{\gamma}\ln(1-\ell^{-\gamma}) + \mathcal{O}\left(N^{-\frac{1}{2}}\right) & ;\ \gamma > 0 \\ (\sqrt{N} - \frac{1}{2})\ln\ell + \mathcal{O}\left(N^{-\frac{1}{2}}\right) & ;\ \gamma = 0 \\ N\ln\ell + \frac{1}{\gamma}\sqrt{N} + \frac{1}{\gamma}\ln\sqrt{N} + \frac{1}{\gamma}\ln(1-\ell^{\gamma}) + \mathcal{O}\left(N^{-\frac{1}{2}}\right) & ;\ \gamma < 0 \end{cases}, \quad \text{(S8)}$$

and thus

$$\Delta S_{\alpha<1}(\gamma) \approx \begin{cases} N\ln\ell & ;\ \gamma \geq 0 \\ \left(\frac{\alpha}{\alpha-1} - \frac{1}{\gamma}\right)\sqrt{N} & ;\ \gamma < 0 \end{cases}, \quad \text{(S9)}$$

which violates Condition **(1)** for $\gamma \neq \frac{\alpha-1}{\alpha}$ and $\alpha < 1$.

As the above examples rule out any $\gamma \neq \frac{\alpha-1}{\alpha}$, we set $\gamma = \frac{\alpha-1}{\alpha}$ and define

$$S^{\text{op}}_\alpha(\rho_A) = \frac{\alpha}{1-\alpha}\ln\left[\sum_n P_n e^{\frac{1-\alpha}{\alpha}S_\alpha(\rho_{A_n})}\right], \quad \text{(S10)}$$

and investigate its properties in the next section.

### PROPERTIES OF $S^{\text{op}}_\alpha$

In this section we prove that Eq. (S10) satisfies conditions **(1)**–**(3)** defined above. It will be convenient to rewrite this measure as defined in the main text:

$$S^{\text{op}}_\alpha = S_\alpha - \frac{1}{1-\alpha^{-1}}\ln\left(\sum_n P_{n,\alpha}^{1/\alpha}\right)$$
$$= S_\alpha - H_{1/\alpha}(\{P_{n,\alpha}\}), \quad \text{(S11)}$$

where

$$P_{n,\alpha} = \frac{\text{Tr}\left(\mathcal{P}_{A_n}\rho_A^\alpha\mathcal{P}_{A_n}\right)}{\text{Tr}\,\rho_A^\alpha} \quad \text{(S12)}$$

$$H_{1/\alpha}(\{P_{n,\alpha}\}) = \frac{1}{1-\alpha}\ln\left(\sum_n P_{n,\alpha}\right). \quad \text{(S13)}$$

We begin by recalling Jensen's famous inequality and a useful property of the $p$-norm.

**Lemma 1** (Jensen's Inequality [S2]). *For a real convex function $f$, the set of $N+1$ real numbers $\{x_0, x_2, \ldots, x_N\}$ in its domain, and the set of non-negative numbers $\{P_0, P_2, \ldots, P_N\}$ such that $\sum_{n=0}^{N} P_n = 1$*

$$\sum_{n=0}^{N} P_n f(x_n) \geq f\left(\sum_{n=0}^{N} P_n x_n\right). \tag{S14}$$

*The inequality is reversed if $f$ is a concave function.*

**Lemma 2** (*p*-norm property). *For a vector $X = (x_0, \ldots, x_{M-1})$ in $\mathbb{R}^M$ with p-norm defined by $\|X\|_p = (\sum_n |x_n|^p)^{p^{-1}}$ then for $p, r \in \mathbb{R}$ with $0 < r \leq p$*

$$\|X\|_p \leq \|X\|_r \leq M^{r^{-1} - p^{-1}} \|X\|_p. \tag{S15}$$

As a direct consequence of Lemma 2:

$$\|X\|_r \geq \|X\|_p \geq M^{p^{-1} - r^{-1}} \|X\|_r. \tag{S16}$$

**Theorem 1.** *For a given reduced density matrix $\rho_A$, $0 \leq \Delta S_\alpha \leq \ln D$ where $D$ is the support of $P_n$.*

*Proof.* Beginning from the definition in Eq. (S10), we write

$$e^{-S_\alpha^{\text{op}}(\rho_A)} = \left[\sum_n P_n \left(\text{Tr}\, \rho_{A_n}^\alpha\right)^{\frac{1}{\alpha}}\right]^{\frac{\alpha}{\alpha-1}} = \left(\|X\|_{\frac{1}{\alpha}}\right)^{\frac{1}{\alpha-1}}, \tag{S17}$$

with $x_n = P_n^\alpha \text{Tr}\, \rho_{A_n}^\alpha$. Also, from Eq. (S2) we can write

$$e^{-S_\alpha(\rho_A)} = (\text{Tr}\, \rho_A^\alpha)^{\frac{1}{\alpha-1}} = \left(\sum_n P_n^\alpha \text{Tr}\, \rho_{A_n}^\alpha\right)^{\frac{1}{\alpha-1}} = (\|X\|_1)^{\frac{1}{\alpha-1}}. \tag{S18}$$

Dividing Eq. (S17) by Eq. (S18) we get

$$e^{\Delta S_\alpha} = \left(\frac{\|X\|_{\frac{1}{\alpha}}}{\|X\|_1}\right)^{\frac{1}{\alpha-1}}. \tag{S19}$$

Using the inequalities (S15) for $\alpha > 1$ and (S16) for $\alpha < 1$ then raising all sides to the exponent $1/(\alpha-1)$ gives

$$1 \leq \left(\frac{\|X\|_{\frac{1}{\alpha}}}{\|X\|_1}\right)^{\frac{1}{\alpha-1}} \leq D, \tag{S20}$$

and thus

$$0 \leq \Delta S_\alpha \leq \ln D. \tag{S21}$$

□

**Theorem 2.** *For any numbers $\alpha, \beta \in \mathbb{R}$ where $\alpha \geq \beta > 0$, $S_\alpha^{\text{op}} \leq S_\beta^{\text{op}}$.*

*Proof.* Consider Jensen's inequality for the convex function $f(x) = x^\delta$ where $\delta \geq 1$:

$$\left(\sum_n P_n x_n\right)^\delta \leq \sum_n P_n x_n^\delta, \tag{S22}$$

The inequality is reversed if $\delta \leq 1$. Now, it is known that the Rényi entropy is a non-increasing function of $\alpha$ and this holds for a fixed number of particles in the subregion so:

$$S_\alpha(\rho_{A_n}) \leq S_\beta(\rho_{A_n}). \tag{S23}$$

First, we consider $\alpha \geq \beta > 1$. Setting $\delta = \frac{\alpha^{-1}-1}{\beta^{-1}-1} \geq 1$ and $x_n = e^{(\beta^{-1}-1)S_\beta(\rho_{A_n})}$ in inequality (S22) we get

$$\left(\sum_n P_n e^{(\beta^{-1}-1)S_\beta(\rho_{A_n})}\right)^{\frac{\alpha^{-1}-1}{\beta^{-1}-1}} \leq \sum_n P_n e^{(\alpha^{-1}-1)S_\beta(\rho_{A_n})}. \tag{S24}$$

Using inequality (S23), for $\alpha > 1$, we can write $e^{(\alpha^{-1}-1)S_\beta(\rho_{A_n})} \leq e^{(\alpha^{-1}-1)S_\alpha(\rho_{A_n})}$ and thus

$$\left(\sum_n P_n e^{(\beta^{-1}-1)S_\beta(\rho_{A_n})}\right)^{\frac{\alpha^{-1}-1}{\beta^{-1}-1}} \leq \sum_n P_n e^{(\alpha^{-1}-1)S_\alpha(\rho_{A_n})}. \tag{S25}$$

Raising both sides of the last inequality to the negative exponent $(\alpha^{-1}-1)^{-1}$ reverses the inequality and taking the logarithm of both sides yields

$$S_\beta^{\text{op}}(\rho_A) \geq S_\alpha^{\text{op}}(\rho_A). \tag{S26}$$

Following the same procedure, it is straightforward to show that the last inequality holds for $1 > \alpha \geq \beta$. $\square$

**Corollary 2.1.** $S_\alpha^{\text{op}}$ *is a lower bound on* $S_1^{\text{op}}$ *for* $\alpha > 1$ *and an upper bound for* $\alpha < 1$.

*Proof.* For $\alpha > 1$ we use Jensen's inequality for the convex function $f(x) = \frac{1}{\alpha^{-1}-1}\ln x$ with $x_n = e^{(\alpha^{-1}-1)S_\alpha(\rho_{A_n})}$. The above inequality is reversed for $\alpha < 1$ as the function $f(x)$ becomes concave. Now

$$S_\alpha^{\text{op}}(\rho_A) = \frac{1}{\alpha^{-1}-1}\ln\left(\sum_n P_n e^{(\alpha^{-1}-1)S_\alpha(\rho_{A_n})}\right) \leq \sum_n P_n S_\alpha(\rho_{A_n}). \tag{S27}$$

By using inequality (S23) once more we get

$$\sum_n P_n S_\alpha(\rho_{A_n}) \leq \sum_n P_n S_1(\rho_{A_n}) = S_1^{\text{op}}(\rho_A), \tag{S28}$$

with the inequality reversed for $\alpha \leq 1$. From (S27) and (S28), we can write

$$S_\alpha^{\text{op}}(\rho_A) \leq \sum_n P_n S_\alpha(\rho_{A_n}) \leq S_1^{\text{op}}(\rho_A) \qquad \alpha \geq 1, \tag{S29}$$

$$S_\alpha^{\text{op}}(\rho_A) \geq \sum_n P_n S_\alpha(\rho_{A_n}) \geq S_1^{\text{op}}(\rho_A) \qquad \alpha \leq 1. \tag{S30}$$

$\square$

**Theorem 3.** $\lim_{\alpha \to 1} S_\alpha^{\text{op}} = S_1^{\text{op}}$.

*Proof.* Taking the limit $\alpha \to 1$ and using Eq. (S11) and (S13) in combination with the fact that $P_{n,1} = P_n$

$$\lim_{\alpha \to 1} S_\alpha^{\text{op}}(\rho_A) = \lim_{\alpha \to 1}\left[S_\alpha - H_{1/\alpha}(\{P_{n,\alpha}\})\right] = S_1 - H_1(\{P_{n,1}\}) = S_1 - H_1 = S_1^{\text{op}}. \tag{S31}$$

$\square$

## FREE FERMIONS

In this section we provide additional details on the asymptotic scaling of the characteristic function $\chi_\alpha(\lambda)$ of the probability distribution $P_{n,\alpha} = \text{Tr}\left[\mathcal{P}_{A_n}\rho_A^\alpha \mathcal{P}_{A_n}\right]/\text{Tr}\,\rho_A^\alpha$ for free fermions in the $d$-dimensional spatial continuum and on a $1d$ lattice.

### $d$-dimensional free fermions in the continuum

Here we provide a detailed derivation of the scaling of the characteristic function for free fermions in the continuum as defined in Eq. (6) of the main text.

Consider the ground state of a $d$-dimensional free gapless Fermi gas, where the Fermi sea is represented by the domain $\Gamma$ in momentum space. Let $\Omega$ be a bounded region in real space and $Q$ be a projection on the region $\Omega$ that is rescaled by a dimensionless factor $\ell$, where $a$ is a fixed short distance. Also, Let $P$ be the projection on the momentum modes in the domain $\Gamma$, where the fermion correlation function is $g(x - x') = \langle x|P|x'\rangle$ and therefore the operator $QPQ$ plays the role of the correlation matrix $\mathsf{C}_A$ in the continuum. For such system the asymptotic scaling of $S_\alpha$ and the logarithm of the characteristic function of $P_n = P_{n,1}$ has been obtained by means of the Widom theorem [S3–S10], where the theorem predicts the asymptotic, $\ell \gg 1$, behavior of $\operatorname{Tr} f(QPQ)$ for class of functions $f$. If $f(t)$ is analytic on the disc $|t| \leq 1$ and $f(0) = 0$, then

$$\operatorname{Tr} f(QPQ) = c_1 f(1)\ell^d + c_2 I(f)\ell^{d-1} \ln \ell + \mathrm{o}\left(\ell^{d-1} \ln \ell\right), \tag{S32}$$

where

$$c_1 = \frac{1}{(2\pi)^d} \int_\Omega \int_\Gamma \mathrm{d}x \, \mathrm{d}p, \qquad c_2 = \frac{1}{(2\pi)^{(d+1)}} \int_{\partial\Omega} \int_{\partial\Gamma} |\boldsymbol{n}_x \cdot \boldsymbol{n}_p| \, \mathrm{d}S_x \, \mathrm{d}S_p, \qquad I(f) = \int_0^1 \mathrm{d}t \, \frac{f(t) - tf(1)}{t(1-t)} \tag{S33}$$

and $\mathrm{o}\left(\ell^{d-1} \ln \ell\right) / \left(\ell^{d-1} \ln \ell\right) \to 0$ as $\ell \to \infty$. Here $\boldsymbol{n}_x$ and $\boldsymbol{n}_p$ are unit vectors normal to $\Omega$ and $\Gamma$, respectively.

Let us now consider the characteristic function of $P_{n,\alpha}$ which we obtain by replacing $\mathsf{C}_A$ with $QPQ$ in:

$$\ln \chi_\alpha(\lambda) = \operatorname{Tr} \ln \left[ \frac{(1 - \mathsf{C}_A)^\alpha + \mathsf{C}_A^\alpha \, \mathrm{e}^{i\lambda}}{(1 - \mathsf{C}_A)^\alpha + \mathsf{C}_A^\alpha} \right]. \tag{S34}$$

By doing so, we get $\ln \chi_\alpha(\lambda) = \operatorname{Tr} f_\alpha(QPQ)$, with $f_\alpha(t) = \ln \left[ \frac{(1-t)^\alpha + t^\alpha e^{i\lambda}}{(1-t)^\alpha + t^\alpha} \right]$ which satisfies Widom theorem conditions for integer $\alpha > 0$, where $f_\alpha(1) = i\lambda$ and

$$I(f_\alpha) = \int_0^1 \mathrm{d}t \, \frac{\ln \left[ \frac{(1-t)^\alpha + t^\alpha e^{i\lambda}}{(1-t)^\alpha + t^\alpha} \right] - i\lambda t}{t(1-t)} = \int_0^1 \mathrm{d}t \, \frac{\ln \left[ \frac{(1-t)^\alpha e^{-i\lambda/2} + t^\alpha e^{i\lambda/2}}{(1-t)^\alpha + t^\alpha} \right]}{t(1-t)}, \tag{S35}$$

where $\lim_{\epsilon \to 0} \int_\epsilon^{1-\epsilon} \mathrm{d}t \, \frac{t - \frac{1}{2}}{t(1-t)} = 0$. Using the substitution $\mu = \frac{t^\alpha}{(1-t)^\alpha + t^\alpha}$, we get

$$I(f_\alpha) = \frac{1}{\alpha} \int_0^1 \mathrm{d}t, \, \frac{\ln \left[ (1-\mu)\mathrm{e}^{-i\lambda/2} + \mu \mathrm{e}^{i\lambda/2} \right]}{\mu(1-\mu)} = \frac{I(f_1)}{\alpha}. \tag{S36}$$

where the integral $I(f_1) = -\lambda^2/2$ [S7]. As a result, we find the result reported in the main text that $\chi_\alpha(\lambda) = \exp\left[ i\lambda c_1 \ell^d - \frac{\lambda^2}{2\alpha} c_2 \ell^{d-1} \ln \ell \right]$ which is to leading order, the characteristic function of the normal distribution

$$P_{n,\alpha} \approx \sqrt{\frac{\alpha}{2\pi c_2 \ell^{d-1} \ln \ell}} \exp\left[ \frac{-\alpha(n - c_1 \ell^d)^2}{2c_2 \ell^{d-1} \ln \ell} \right]. \tag{S37}$$

### 1-dimensional free fermions on a lattice

All source code and data related to our calculations of $1d$ lattice fermions can be found online [S11]. Starting from the discrete characteristic function $\chi_\alpha(\lambda)$ of $P_{n,\alpha}$ defined in the main text

$$\ln \chi_\alpha(\lambda) = \sum_{j=1}^\ell \ln \left[ 1 - \nu_{j,\alpha} + \nu_{j,\alpha} \, \mathrm{e}^{i\lambda} \right], \tag{S38}$$

and using the asymptotic ($\ell \gg 1$) form for the eigenvalues $\nu_{j,\alpha}$ [S12–S14]

$$\nu_{j,\alpha} = \left[ 1 + \exp\left( \frac{-\alpha \pi^2 (\ell - 2j + 1)}{2[\ln(8\ell) + \gamma_{\mathrm{em}}]} \right) \right]^{-1}, \tag{S39}$$

we evaluate the asymptotic behavior of $\chi_\alpha(\lambda)$ by replacing the summation over the index $j$ by an integration over the variable $k \in (-\ell/(2\ln\ell), \ell/(2\ln\ell))$ with $-(\ell - 2j + 1)/2 \to k\ln\ell$. For $\ell \gg 1$ ($\ln\ell \gg \ln 8 + \gamma_{\rm em}$) we can write

$$\ln \chi_\alpha(\lambda) \approx \ln \ell \int_{-\ell/(2\ln\ell)}^{\ell/(2\ln\ell)} {\rm d}k \, \ln\left[1 - \nu_\alpha(k) + \nu_\alpha(k) \, {\rm e}^{i\lambda}\right] \tag{S40}$$

$$\nu_\alpha(k) = \left[1 + \exp\left(\alpha\pi^2 k\right)\right]^{-1} . \tag{S41}$$

It follows that ${\rm d}k = -\frac{1}{\alpha\pi^2}\frac{d\nu_\alpha}{\nu_\alpha(1-\nu_\alpha)}$. By pulling out a factor ${\rm e}^{i\lambda/2}$ from Eq. (S40) and performing another change of variables from $k \to \nu$ we get

$$\ln \chi_\alpha(\lambda) \approx \frac{\ln(\ell)}{\alpha\pi^2} \int_{\nu_\alpha(+\ell/(2\ln\ell))}^{\nu_\alpha(-\ell/(2\ln\ell))} {\rm d}\nu_\alpha \frac{\ln\left[(1 - \nu_\alpha(k))\, {\rm e}^{-i\lambda/2} + \nu_\alpha(k)\, {\rm e}^{i\lambda/2}\right]}{\nu_\alpha(1-\nu_\alpha)} + \ln(\ell) \int_{-\ell/(2\ln\ell)}^{\ell/(2\ln\ell)} {\rm d}k\, \frac{i\lambda}{2} = \frac{\ln(\ell)}{\alpha\pi^2}\mathcal{I}_\alpha(\ell) + i\lambda\frac{\ell}{2} . \tag{S42}$$

In the limit $\ell \to \infty$, $\nu_\alpha(\ell/(2\ln\ell)) \to 0$ and $\nu_\alpha(-\ell/(2\ln\ell)) \to 1$ leading to

$$\lim_{\ell \to \infty} \mathcal{I}_\alpha(\ell) = \int_0^1 {\rm d}\nu_\alpha \frac{\ln\left[(1-\nu_\alpha(k))\,{\rm e}^{-i\lambda/2} + \nu_\alpha(k)\,{\rm e}^{i\lambda/2}\right]}{\nu_\alpha(1-\nu_\alpha)} = -\frac{\lambda^2}{2} \tag{S43}$$

thus

$$\ln \chi_\alpha(\lambda) \approx i\lambda \frac{\ell}{2} - \frac{\lambda^2}{2}\left(\frac{\ln(\ell)}{\alpha\pi^2}\right). \tag{S44}$$

Eq. (S44) describes the characteristic function of a normal distribution with average $\langle n \rangle = \frac{\ell}{2}$ and variance $\sigma_\alpha = \frac{\ln(\ell)}{\alpha\pi^2}$ as reported in the main text below Eq. (9).

---



\end{document}